\documentclass[12pt]{iopart}

\usepackage{lineno}
\usepackage{amstext}
\usepackage{graphicx}
\usepackage{url}
\usepackage{nameref}
\usepackage{natbib}
\setcitestyle{authoryear}

\begin{document}

\title[Convective environments within Mediterranean cyclones]{Convective environments within Mediterranean cyclones}

\author{Alice Portal$^{1}$\footnote{Present address: Institute of Atmospheric Sciences and Climate (CNR-ISAC), National Research Council of Italy, Bologna, Italy}, Andrea Angelidou$^{1}$, Raphael Rousseau-Rizzi$^{2}$, Shira Raveh-Rubin$^{3}$, Yonatan Givon$^{3}$, Jennifer L Catto$^{4}$, Francesco Battaglioli$^{5,6}$, Mateusz Taszarek$^{7,8}$, Emmanouil Flaounas$^{9}$, Olivia Martius$^{1}$}

\address{$^{1}$ Institute of Geography, Oeschger Centre for Climate Change Research, University of Bern, Bern, Switzerland}
\address{$^{2}$ Centre de Recherche d'Hydro-Québec, Varennes, Quebec, Canada}
\address{$^{3}$ Department of Earth and Planetary Sciences, Weizmann Institute of Science, Rehovot, Israel}
\address{$^{4}$ Department of Mathematics and Statistics, University of Exeter, Exeter, United Kingdom}
\address{$^{5}$ European Severe Storms Laboratory (ESSL), Wessling, Germany}
\address{$^{6}$ Institut für Meteorologie, Freie Universität Berlin, Berlin, Germany}
\address{$^{7}$ Department of Meteorology and Climatology, Adam Mickiewicz University in Poznań, Poland}
\address{$^{8}$ Cooperative Institute for Severe and High-Impact Weather Research \& Operations, University of Oklahoma, Norman, Oklahoma, United States}
\address{$^{9}$ Institute of Oceanography, Hellenic Centre for Marine Research, Athens, Greece}
\ead{a.portal@isac.cnr.it}
\vspace{12pt}

\begin{abstract}
Understanding convective processes leading to severe weather hazards within Mediterranean cyclones is relevant for operational forecasters, insurance industry, and enhancing societal preparedness.
In this work we examine the climatological link between Mediterranean cyclones and atmospheric conditions conducive to the formation of severe convection and convective hazards (convective precipitation, lightning and hail potential). Using ATDnet lightning detections we find that, from autumn to spring, 20 to 60\% of lightning hours over the Mediterranean basin and adjacent land regions are associated with the presence of a nearby cyclone. Based on reanalysis data, severe convective environments, deep, moist convection (i.e., lightning potential) and related hazards are frequent in the warm sector of Mediterranean cyclones and to the north-east of their centres. In agreement with previous literature, convective processes and hazards peak approximately six hours prior to the time of minimum pressure of the cyclone centre.
Moreover, severe convective environments are often detected in cyclone categories typical of transition seasons (especially autumn) and summer, while they are rarer in deep baroclinic cyclones with peak occurrence during winter.
Finally, we show that dynamical cyclone features distinguish regions favourable to deep, moist convection. Warm conveyor belts of Mediterranean cyclones, characterised by large-scale ascent and located in regions of high thermodynamic instability, have the largest lightning potential. The potential is only half as intense along the cyclones' cold fronts.
\end{abstract}

%
\vspace{2pc}

\ack
AP is grateful for the many instructive chats about convection with Monika Feldmann.
The research contributes to the efforts of COST Action CA19109 under MedCyclones -- European network for Mediterranean cyclones in weather and climate.

\section*{Funding Statement} 
SRR and YG acknowledge funding from the Israel Science Foundation (grant number 1242/23) and the De Botton Center for Marine Science, Weizmann Institute. MT acknowledges funding from the Polish National Science Centre (grant number 2020/39/D/ST10/00768).
AP and ORM acknowledge support from the Swiss National Science Foundation Grant Number IZCOZ0\_205461 and from the nextGEMS Project funded by the European Union's Horizon 2020 research and innovation program under Grant agreement numbers 821205 and 101003470. 
This work was made possible by the TROPICANA program of the Institut Pascal at Université Paris-Saclay with the support of the program “Investissements d’avenir” ANR-11-IDEX-0003-01.

\section*{Data Availability Statement}
ERA5 reanalysis fields are publicly available at https://doi.org/10.24381/cds.adbb2d47; ERA5 post-processed data of lightning and hail probability can be made available on request from author Dr Francesco Battaglioli (francesco.battaglioli@essl.org). ATDnet lightning observations were provided by the UK Met Office. Mediterranean cyclone tracks are provided as a supplement of \citep{flaounas2023composite}, while the Mediterranean cyclone impact area dataset is openly available in zenodo at http://doi.org/10.5281/zenodo.13758677.

\submitto{\ERL}

\maketitle
%
%

\section{Introduction}
\label{sec:intro}

Mediterranean weather and climate are influenced by cyclonic activity \citep{campins2011climatology}. 
Mediterranean cyclones (henceforth MedCys), as extratropical cyclones in general, develop from a disturbance in a baroclinically unstable environment \citep{fita2006intercomparison,flaounas2015dynamical,flaounas2022mediterranean}.
Although generally weaker, smaller and shorter-lived than Atlantic extratropical cyclones \citep{campins2011climatology,lionello2016objective,corner2024classification}, MedCys can induce extreme and damaging weather conditions \citep[e.g.,][]{pfahl2014characterising,nissen2010cyclones,portal2024linking,michaelides2018reviews}.

Impactful weather is linked to the character of the cyclone's circulation, but also depends on the presence of (smaller-scale) convective phenomena embedded within the cyclone structure. The phenomena range in scale and severity, from embedded cellular convection \citep{fuhrer2005embedded} and small-scale thunderstorms to fully developed mesoscale convective systems \citep[e.g.][]{schumacher2020formation}.
In general, these enhance precipitation rates and surface wind gusts, and, in the case of strong supercells, can produce hail and tornadoes \citep{oertel2019convective,oertel2021observations,flaounas2016processes, flaounas2018heavy,tochimoto2021characteristics}.

Environmental parameters useful for identifying severe convection potential include convective available potential energy (henceforth CAPE) and vertical wind shear. CAPE combines temperature lapse rate and moisture to quantify thermodynamic instability, while vertical shear promotes convective organisation \citep{puvcik2015proximity,taszarek2020severeB}. In this work we distinguish ``deep, moist convection'' \citep[DMC,][]{doswell2001severe}, i.e., convective activity diagnosed through lightning, from ``severe convective environments'' (SCEs), i.e., reanalysis environmental conditions favourable to the development of severe convection. We also discuss ``shallow or elevated convection'', typically embedded in regions of stratiform precipitation and characterised by limited CAPE and forced ascent, such as in cyclones' warm conveyor belts \citep{herzegh1980mesoscale,oertel2019convective} and flow over orography \citep{fuhrer2005embedded}. 
 
Convective activity is expected in the extratropical cyclone sectors with significant CAPE and forced ascent, i.e., close to the cyclone's low-pressure centre \citep{tracton1973role,flaounas2018heavy,galanaki2016lightning}, in the warm conveyor belt \citep{flaounas2018heavy,oertel2019convective,rasp2016convective,blanchard2020organization, herzegh1980mesoscale}, along cold frontal bands \citep{naud2015cloudsat,raveh2016large} and where the flow impinges upon orography and coastal boundaries \citep{lionello2006cyclones,tochimoto2021characteristics}.  The relatively high Mediterranean Sea surface temperatures support convection by providing heat and humidity to the atmosphere \citep[e.g.,][]{rigo2019improved}, particularly in autumn when the air aloft is colder than the surface and cyclones are frequent \citep{flaounas2019heavy,lombardo2024climatology}.
The importance of convective phenomena around MedCys has been recognised by the scientific community studying Mediterranean weather \citep[e.g.,][]{flaounas2022mediterranean}, in particular by \cite{galanaki2016lightning} and \cite{flaounas2018heavy} relating DMC and MedCys, and by \cite{tochimoto2021characteristics} investigating MedCys associated with tornado development over Italy.

One of the challenges specific to MedCys is the large variety of systems concerned.
\cite{givon2024process} demonstrated how a separation of MedCys into nine clusters based on differences in the upper-level flow discloses meaningful intra-cluster commonalities. The commonalities emerge in the synoptic, seasonal and geographic setting, and extend to the surface footprints \citep{rousseau2023stormrelative}. 
After a preliminary assessment of the seasonal association between observed DMC (i.e., lightning detections) and MedCys (Section~\ref{ssec:Med}), we address the following questions:
\begin{enumerate}
    \item Is the presence of SCEs specific to the upper-level clusters of \cite{givon2024process}?  (Section~\ref{ssec:Cl})
    \item How are the spatial patterns and temporal evolution of convective environments and hazards organised by MedCys? (Section~\ref{ssec:spatdistr}) And how does the spatial arrangement of DMC (i.e., lightning potential) depend on the presence of cold fronts and warm conveyor belts around MedCys? (Section~\ref{ssec:DF})
\end{enumerate}

\section{Data and Methods}
\label{sec:met}

We diagnose DMC using ATDnet detections of lightning strikes \citep{enno2020lightning}, from which we extract lightning hours (hours with $\geq1$ strike). The measurements cover 34$^\circ$--71.25$^\circ$N and 13$^\circ$W--48$^\circ$E, excluding the south-eastern Mediterranean basin, with a resolution of 0.25$^\circ$x0.25$^\circ$ in years 2008--2020. 

From ERA5 reanalysis \citep{hersbach2020era5}, interpolated to a resolution of 0.5$^\circ$x0.5$^\circ$ over years 1980-2020, 
we analyse convective parameters such as most-unstable CAPE, vertical wind shear between 500~hPa ($\sim$6~km) and 10~m above ground ($\sim$0~km) (WS06) and 6h accumulation of convective precipitation (CP). We label CP moderate-accumulation CP events when above 5$\,$mm/6h, while we identify SCEs where the variable WMAXSHEAR$\,\equiv \sqrt{2*\text{CAPE}}*\text{WS06}$ is above 500 m$^2$s$^{-2}$ \citep{taszarek2020severeB}. All fields are analysed at a 6h frequency (time steps 00, 06, 12 and 18 UTC).

The IFS model, producing ERA5 data, comprises deep, shallow or mid-level convective parametrisation schemes depending on the base-height and depth of the updraft \citep[see documentation in][]{IFSdocumentation2016}. Note that the shallow and mid-level schemes can be relevant for representing weak (elevated) convection in cyclones' warm conveyor belts \citep[e.g.,][]{oertel2019convective}. ERA5 underestimates CAPE, specially in SCEs \citep{taszarek2021comparison}, and displays negative (positive) biases for heavy (light) CP events \citep{lavers2022evaluation}. Overall, the convective-over-total precipitation is overestimated in the Mediterranean region \citep{lombardo2024climatology}. 
Despite this, the consistency of climatological and event patterns of (convective) precipitation with observed estimates \citep{lombardo2024climatology,owen2021compound} and the enhanced performance compared to previous products \citep{taszarek2021comparison}, provide some confidence on IFS convective parametrisation, especially for the purpose of climatological studies.

The probability of lightning and hail are derived from ERA5 hourly data, according to the Additive Regressive Convective Hazard Model (AR-CHaMo) \citep{battaglioli2023modeled}. The model, trained using observations and ERA5 predictors, assigns hourly probabilities per grid-point and time.
A comparison between AR-CHaMo ERA5 and ATDnet lightning around MedCys evidences the good performance of AR-CHaMo (Figure~S1). 

\label{ssec:cl-gr}
\begin{figure}[h]
\centering
\includegraphics[width=.8\textwidth]{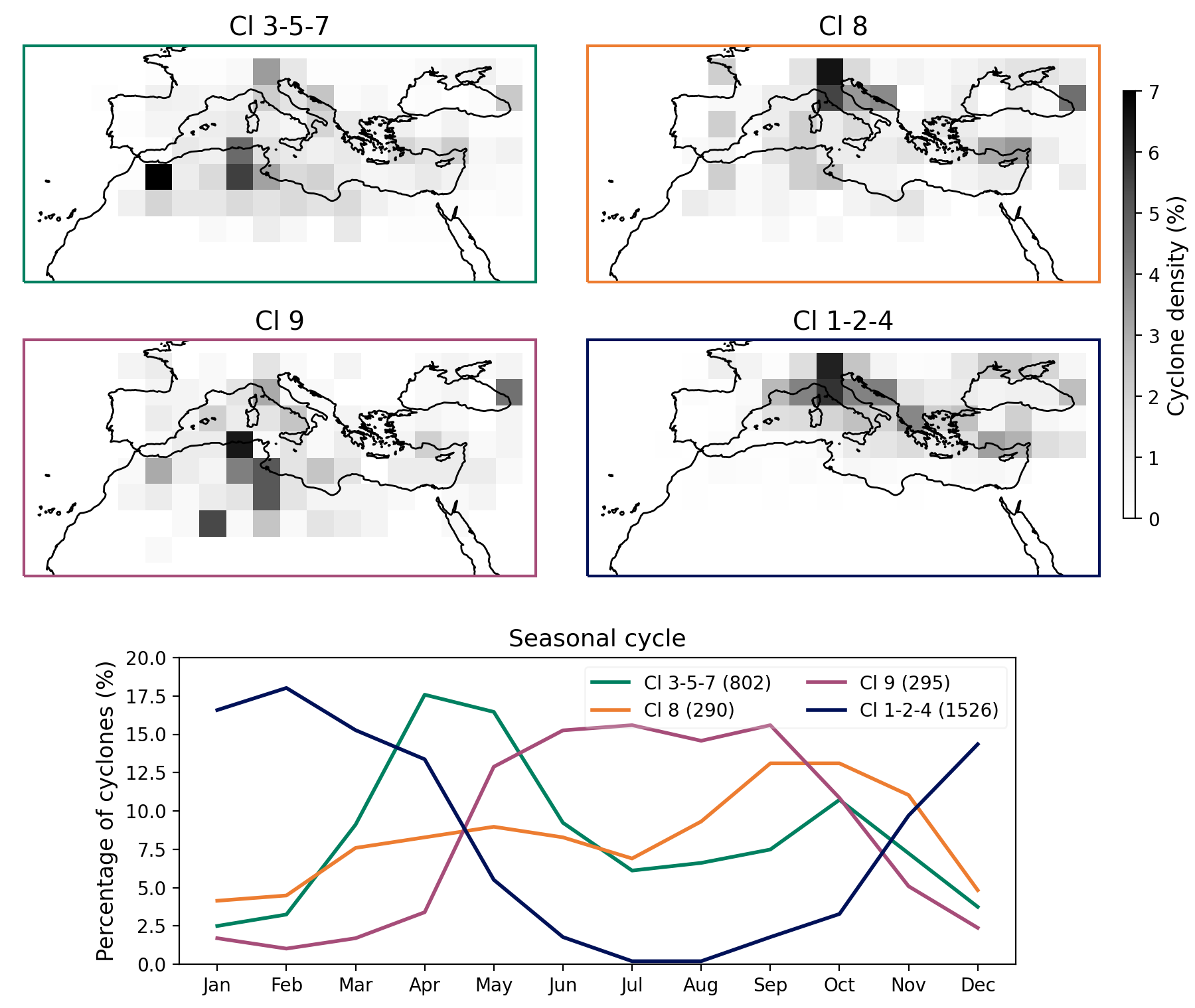}
\caption{Spatial and temporal distribution of MedCy tracks crossing the extended Mediterranean region 20$^\circ$--50$^\circ$N and 20$^\circ$~W--45$^\circ$~E by cluster group. The position and time of the cyclone maximum intensity are considered. The total number of cyclones per cluster group is represented by the value in parenthesis within the legend}
\label{fig:distr}
\end{figure}

We retrieve MedCy tracks from the composites of \cite{flaounas2023composite} selected in the Mediterranean region 20$^\circ$--50$^\circ$N and 20$^\circ$~W--45$^\circ$~E, choosing tracks detected by at least 5 out of 10 tracking methods (CL~5). 
The tracks are separated into nine clusters based on the upper-level potential vorticity (PV) pattern, following \cite{givon2024process}. These are then joined in four groups representative of the convective characteristics of each cluster (see inter-cluster comparison in Figures~\ref{fig:wmxshear_freq},~S2). The cluster groups follow a specific geographical and seasonal frequency (Figure~\ref{fig:distr}):
\begin{itemize}
\item Cl~3-5-7 (cut-off, daughter and anticyclonic wave-breaking lows) are most frequent in the transition seasons (primarily spring) in the southern Mediterranean region;
\item Cl~8 (cyclonic wave-breaking lows) are transition-season (primarily autumn) cyclones which occur mostly in the northern Mediterranean region;
\item Cl~9 (short-wave cut-off lows) are summer cyclones which typically occur in the southern Mediterranean and North-West Africa;
\item Cl~1-2-4 (stage-A/B lee-lows and anticyclonic/cyclonic wave-breaking lows) are winter baroclinic cyclones which mainly occur in the northern Mediterranean.
\end{itemize}
Cl~6 (heat~lows) is omitted because of its dry (land-based) development and negligeable moist processes. 

Composites are computed at the time of cyclone maximum intensity (sea-level pressure minimum along the track), which is also used as reference time ``0h'' in time evolution analyses.
Time series are computed as the cluster-group average of a 200~km radius spatial mean around the cyclone centre. The averages are considered significant if they fall outside the [5th,95th] percentile of a distribution of 2000 Monte-Carlo sample averages. Each random sample is built to reproduce the same geographical and seasonal distribution as the original sample; this is done by re-allocating each cyclone centre of a cluster group to a random year (1980--2020) and a random calendar day within [-5,+5] from the true day. We account for false discovery rate with control level $\alpha=0.05$ \citep{wilks2011statistical}.

Cyclone features corresponding to cold fronts (CFs, the 2.5$^\circ$ extension of CF lines as in \cite{sansom2024objective}) and warm conveyor belts (WCBs, \cite{madonna2014warm,heitmann2023warm}) are considered. WCBs are distinguished in (potentially overlapping) low-level inflow and mid-level ascent region, based on the presence of WCB Lagrangian trajectories below 800~hPa and between 800~and 400~hPa, respectively. 
Moreover, a cyclone impact area IA, defined as a 500~km radius around the centre \citep[consistent with][]{miglietta2013analysis} extended by CF and WCB objects intercepting this radius, identifies the area of influence of a cyclone.  
Cyclone features and IA objects are available from 1980 to 2019 at \url{https://doi.org/10.5281/zenodo.13758677}.

\section{Results}
\label{sec:res}

\subsection{Deep, moist convection (DMC) in the Mediterranean region}
\label{ssec:Med}

We relate lightning detections (proxy of DMC) to MedCy tracks over the climatological period when both datasets are available (2008--2019) at a 6h frequency. Starting from the analysis of DMC seasonality, we support the relevance of MedCys for regional DMC and extend results by \cite{galanaki2016lightning} . 

\begin{figure}[h]
\centering
\includegraphics[width=\textwidth]{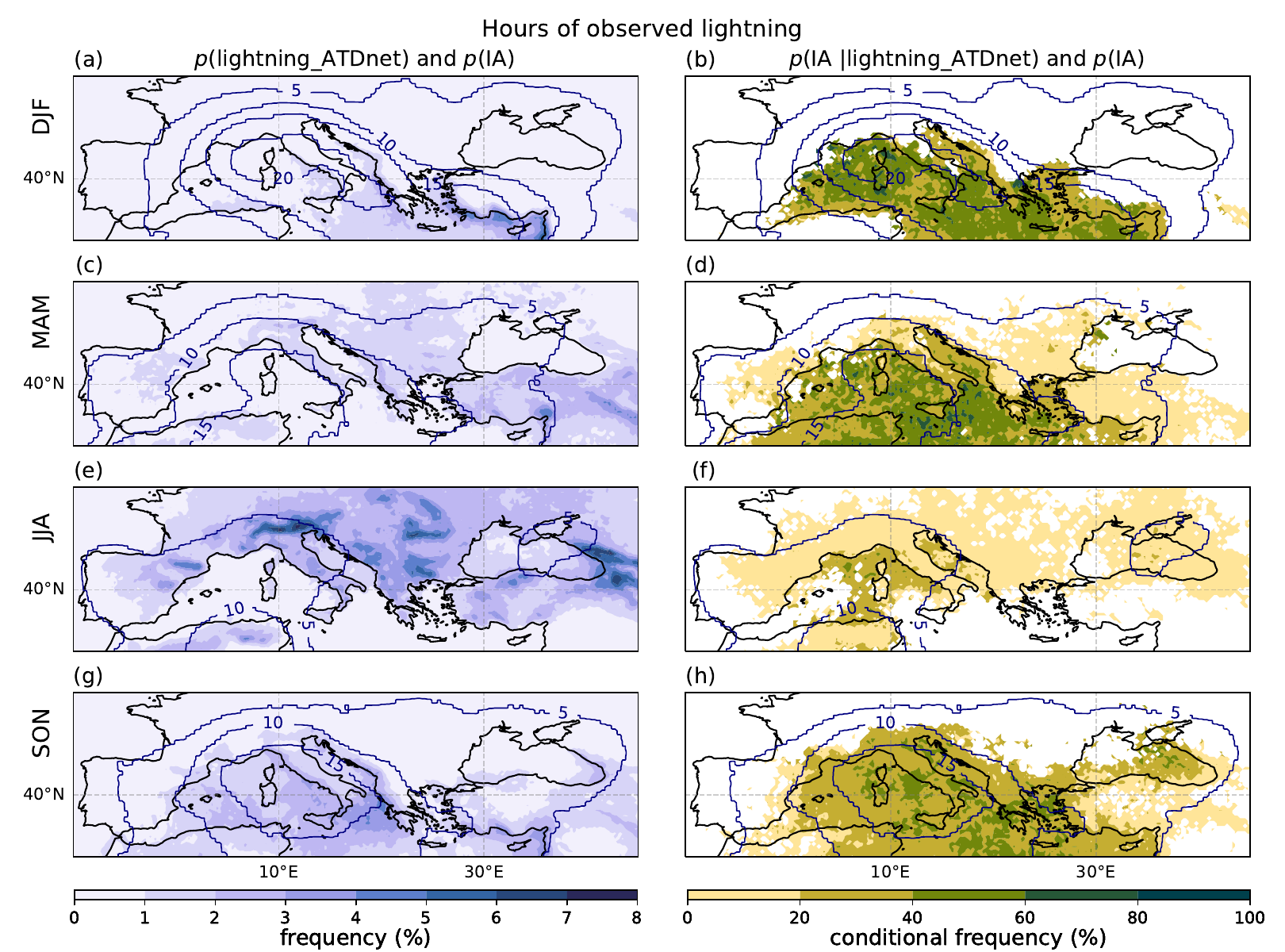}
\caption{Panels (a,c,e,g) show the frequency of lightning hours per season (shading). Panels (b,d,f,h) show the percentage of lightning hours happening within a MedCys' impact area (IA) per season (shading). Grid points where fewer than four (IA~$\wedge$~lightning\_ATDnet) events are detected in the respective season over the observational period (2008--2019) are masked in white. Contours represent the seasonal frequency of MedCy IA}
\label{fig:light_cycl_freq}
\end{figure}

The frequency of lightning hours shows a distinct seasonality in intensity and location (Figure~\ref{fig:light_cycl_freq}(a,c,e,g)). While in spring and summer lightning is most frequent over land  (Figure~\ref{fig:light_cycl_freq}(c),(e)), in autumn and winter the majority of lightning hours is over the sea and coastal areas facing southwest (Figure~\ref{fig:light_cycl_freq}(g),(a)). The most active lightning areas in summer are orography to the north of the Mediterranean basin and around the Black Sea, in autumn the central Mediterranean Sea and its coastal regions, in winter the coasts of Turkey, Syria and Lebanon. Our results match with previous estimates of the annual cycle in convective activity over the Mediterranean region \citep[e.g.,][]{lombardo2024climatology,taszarek2019climatology,taszarek2020severeA, manzato2022pan,galanaki2018thunderstorm}.

The percentage of lightning hours co-occurring with MedCys, i.e., within a cyclone's impact area, also varies by season and region (Figure~\ref{fig:light_cycl_freq}(b,d,f,h)). Summer DMC shows a weak association with MedCys  (Figure~\ref{fig:light_cycl_freq}(f)) because other processes related to the diurnal radiative surface heating trigger DMC over land and orography \citep[e.g.,][]{behrendt2011observation,dolores2019influence}. Indeed, in some orographic regions such as the southern Alps, thunderstorms can develop regardless of the tropospheric flow \citep{ghasemifard2024changing}. In winter, spring, and autumn 20\% to 60\% of the lightning activity over the sea is associated with a nearby MedCy (Figure~\ref{fig:light_cycl_freq}(b,d,h)). The link with MedCys peaks in the western Mediterranean in winter, and in the central Mediterranean in spring and autumn. In autumn the percentage of lightning co-occurring with MedCys is smaller, but extends further inland compared to other seasons (e.g., Northern Italy and Balcanic peninsula). In spring Northern Algeria and Tunisia experience a high proportion of cyclone-related lightning activity, related to the southward extension of the MedCy tracks (Figure~\ref{fig:light_cycl_freq}(d)).

Finally, the frequency of lightning hours when a cyclone is present increases by a factor greater compared to to times with no cyclones (Figure~S4). Although we verify the existence of a strong link between MedCys and lightning hours over the Mediterranean Sea and in adjacent land regions, our values constitute a lower bound because of the exclusion of lower-confidence cyclone tracks which nevertheless contribute to DMC.

\subsection{Convective environments associated with MedCys}
\label{ssec:Cl}

We investigate here where and when to find convective environments and hazards around a cyclone centre. MedCys, taken at the time of their maximum intensity, are first separated into the nine clusters of \cite{givon2024process}, and subsequently shown in cluster groups representative of the individual clusters' characteristics.

\begin{figure}[h]
\centering
\includegraphics[width=.8\textwidth]{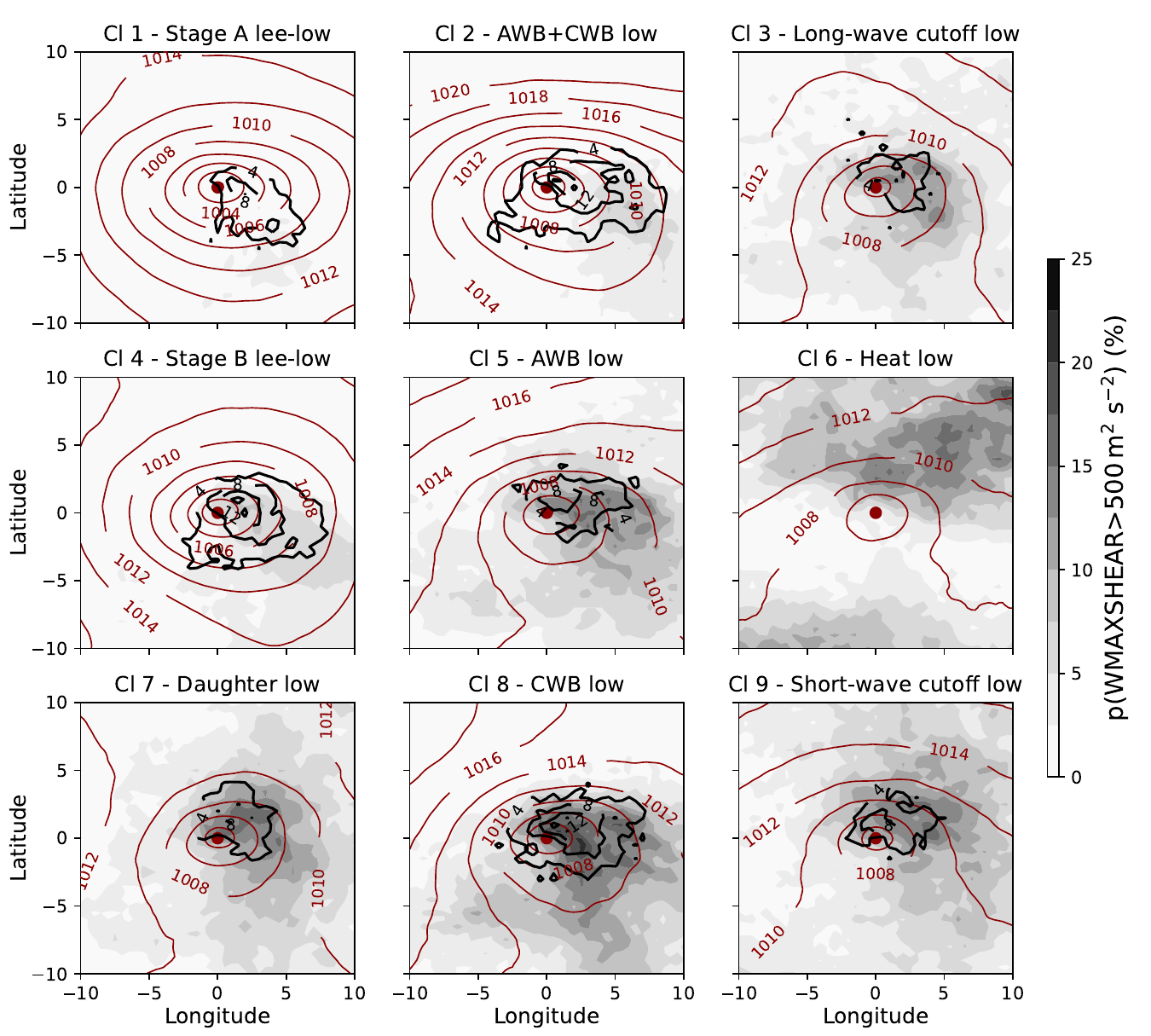}
\caption{Composite showing the percentage frequency of SCEs (WMAXSHEAR$>500\,$m$^2$s$^{-2}$, shading), and of intense 6h CP (black contours indicating CP$>5\,$mm/6h at 4,~8,~12,~16\% frequency) and the average sea level pressure (dark red contours, hPa) per cluster, centred at the cyclone centre at the time of maximum cyclone intensity}
\label{fig:wmxshear_freq}
\end{figure}

The frequency of SCEs (WMAXSHEAR$>500\,$m$^2$s$^{-2}$) and of moderate-accumulation CP events (CP$>5\,$mm/6h) are displayed in Figure~\ref{fig:wmxshear_freq}. While CP events identify areas of active convection (with no specification on depth and severity), high values of WMAXSHEAR are relevant to detect environments favorable to the development of severe thunderstorms and convective hazards \citep[e.g., large hail, tornadoes and severe winds; see][]{taszarek2020severeB}. Cl~8 displays the highest SCE frequency peaking to the north of the centre and in the (south-eastern) warm sector. It also displays amongst the highest occurrences of moderate CP. Cl~3,~5,~7 and~9 show a similar spatial distribution in SCEs, although the frequencies of above-threshold WMAXSHEAR and CP are smaller.
Moderate-accumulation CP is frequent in Cl~1,~2 and~4, which however display the lowest SCE frequency -- above-threshold WMAXSHEAR occurrs only 2.5\% of the time steps.
On the contrary, in Cl~6 SCEs are frequent to the north of the cyclone centre, but, because above-threshold CP is rare (below 4\%) it is neglected in the following.
The results of the frequency analysis reflect in the average WMAXSHEAR and 6h-accumulated CP by cluster (Figure~S2).

\subsubsection{Spatial and temporal characteristics of convective environments around MedCys\\}
\label{ssec:spatdistr}
\begin{figure}[h]
\centering
\includegraphics[width=.8\textwidth]{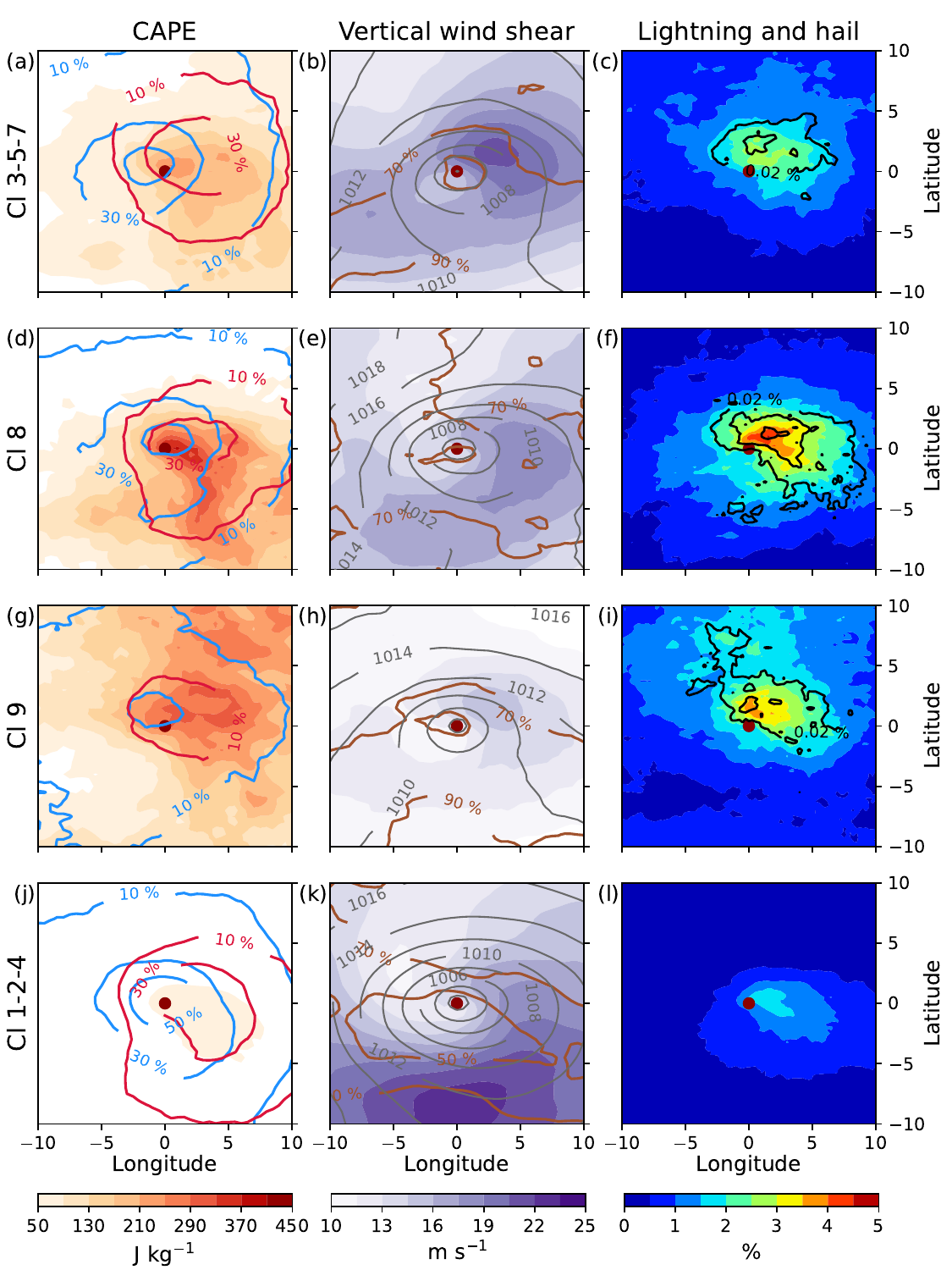}
\caption{Composites of (a,d,g,j) CAPE , (b,e,h,k) vertical wind shear, and (c,f,i,l) modelled AR-CHaMo lightning probability, by cluster group. Contours indicate cold-front frequency (blue, \%), warm-conveyor-belt frequency (red, \%), land fraction (brown, \%), mean sea-level pressure (grey, hPa), modelled AR-CHaMo hail probability (black, at 0.02,~0.04,~0.06\% levels)}
\label{fig:composites}
\end{figure}
In this Section we examine CAPE and wind shear, components of WMAXSHEAR, together with modelled AR\-CHaMo lightning and hail probability. 
In Figure~\ref{fig:composites} the difference between highly and moderately convective transition-season cyclones (Cl~8 vs Cl~3-5-7) emerges in the pronounced thermodynamic instability of Cl~8 (cf. Figure~\ref{fig:composites}(d) and~(a)). It is linked to the more common occurrence of Cl~8 above sea surface compared to Cl~3-5-7 (cf. brown contours in Figure~\ref{fig:composites}(e) and (b)) and to its higher autumn frequency (Figure~\ref{fig:distr}). In fact, the presence of warm autumn SSTs relative to the temperature of the atmosphere favours the development of high CAPE environments. The importance of warm SST in enhancing thunderstorm severity was also highlighted by \cite{gonzalez2023anthropogenic} in the context of the derecho of 17 August 2022 over the central Mediterranean. 
The spatial distribution of CAPE and of lightning and hail probability in Cl~8 and~3-5-7 peaks in the warm sector of the cyclone and shows a substantial overlap with WCBs and high values of wind shear (Figure~\ref{fig:composites}(a-f)).
Although Cl~8 displays weaker wind shear than Cl~3-5-7 (cf. Figure~\ref{fig:composites}(e) and~(b)), its higher CAPE explains the larger lightning and hail probabilities (cf. Figure~\ref{fig:composites}(f) and~(c)). 

In Cl~9, comprising mainly summer cyclones, high CAPE values are accompanied by weaker wind shear and lower frequency of cyclone features compared to transition season clusters (Figure~\ref{fig:composites}(g,h)). Although the lightning probability is higher than in Cl~3-5-7, the hail probability exhibits comparable values --- potentially suppressed by the weaker wind shear (cf. Figure~\ref{fig:composites}(h,i) and (b,c)). As indicated in \cite{kumjian2020hail} and \cite{nixon2023hodographs}, hail develops with the right balance between high CAPE and sufficient wind shear. 

The presence of severe convection in winter baroclinic Cl~1-2-4 is limited by the weak and infrequent CAPE (Figure~\ref{fig:composites}(j)). Nonetheless, the presence of abundant CP accumulations (Figure~S2) suggests the presence of shallow or weak elevated convection before CFs or within WCBs (Figure~\ref{fig:composites}(j)), supported by the frequent occurrence of these cyclones over the sea (Figure~\ref{fig:composites}(k)). We note that large-scale precipitation, associated with frequent WCBs and cold fronts, is highest in this cluster group \citep{givon2024process} and possibly constitutes a limiting factor for the growth of strong thermodynamic instability.

\begin{figure}[h]
\centering
\includegraphics[width=
\textwidth]{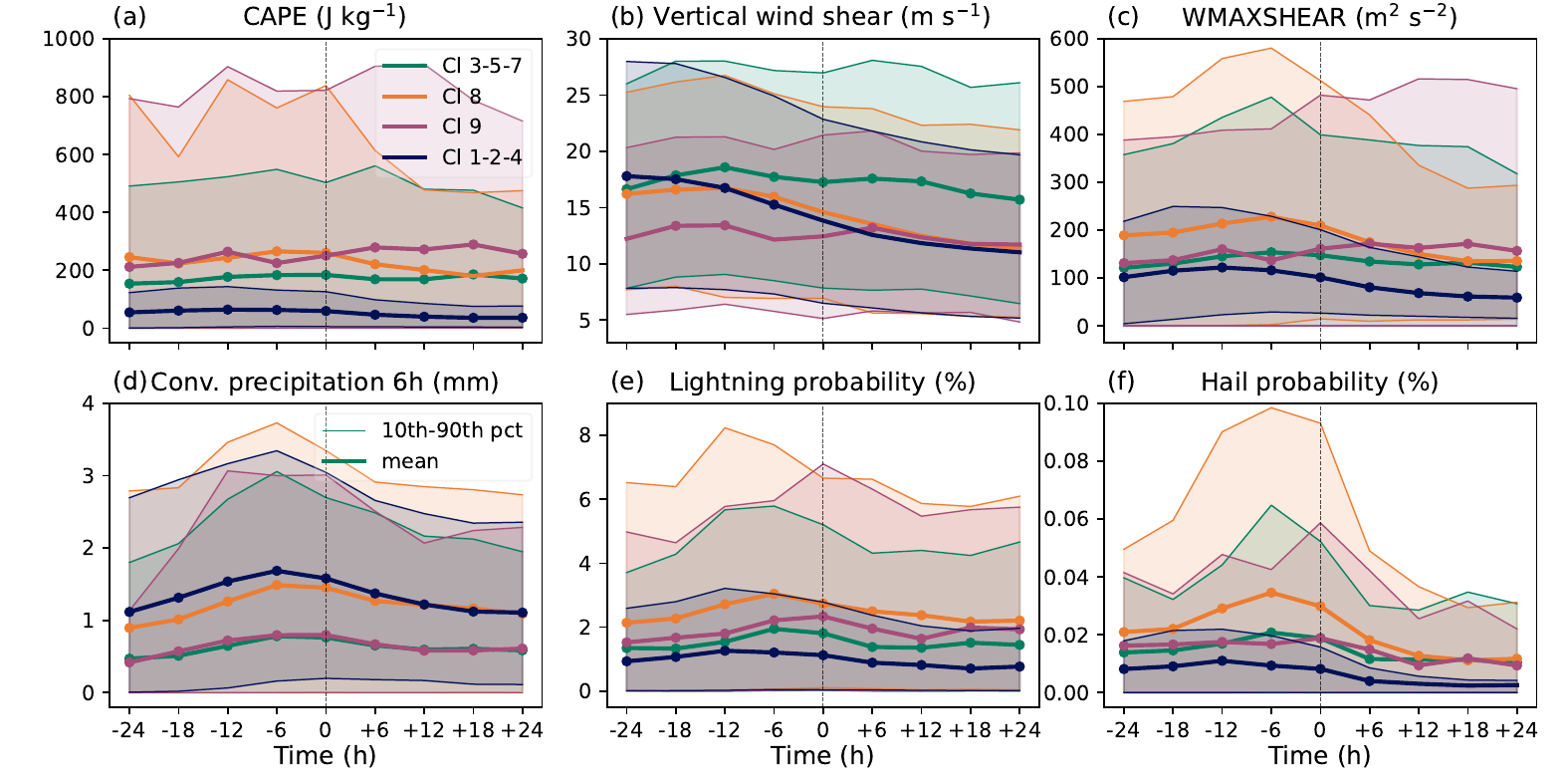}
\caption{Time series of (a) CAPE, (b) vertical wind shear, (c) WMAXSHEAR, (d) 6h-accumulation of CP, (e) lightning probability, and (f) hail probability in the 200 km radius around the cyclone centre. For every cluster group the thick line indicates the mean value, with significant values identified by a full circle, and the thin lines indicate the 10th and 90th percentiles across the cyclones. Time steps are centred around the time of cyclone maximum intensity (vertical dashed line)}
\label{fig:tseries}
\end{figure}

The time evolution of convective variables is analysed by averaging in a 200~km radius around the cyclone centre from -24h to +24h relative to the time of cyclone maximum intensity. The values of CAPE are rather constant in time (Figure~\ref{fig:tseries}(a)), although the upper decile displays a strong variability and reaches its maximum in the cyclone deepening stage. Vertical wind shear generally decreases in time, as the cyclonic circulation depletes the environmental baroclinic instability (Figure~\ref{fig:tseries}(b)). Severe thunderstorm potential (WMAXSHEAR), lightning and hail probability tend to peak before time ``0h'' and prior to 6h-accumulated CP (Figure~\ref{fig:tseries}(c-f)), in accordance with \cite{galanaki2016lightning,flaounas2018heavy}. Hail decays more rapidly than lightning probability. 
The time series also highlight differences across clusters groups, with highest CAPE and lightning and hail probability in transition-season and summer clusters (Figure~\ref{fig:tseries}(a,e,f)). 
Once more, 6h-accumulated CP is abundant in winter cyclones (Cl~1,~2,~4) and in Cl~8 (Figure~\ref{fig:tseries}(d)).

\subsection{Lightning probability and cyclone features}
\label{ssec:DF}
\begin{figure}[h]
\centering
\includegraphics[width=\textwidth]{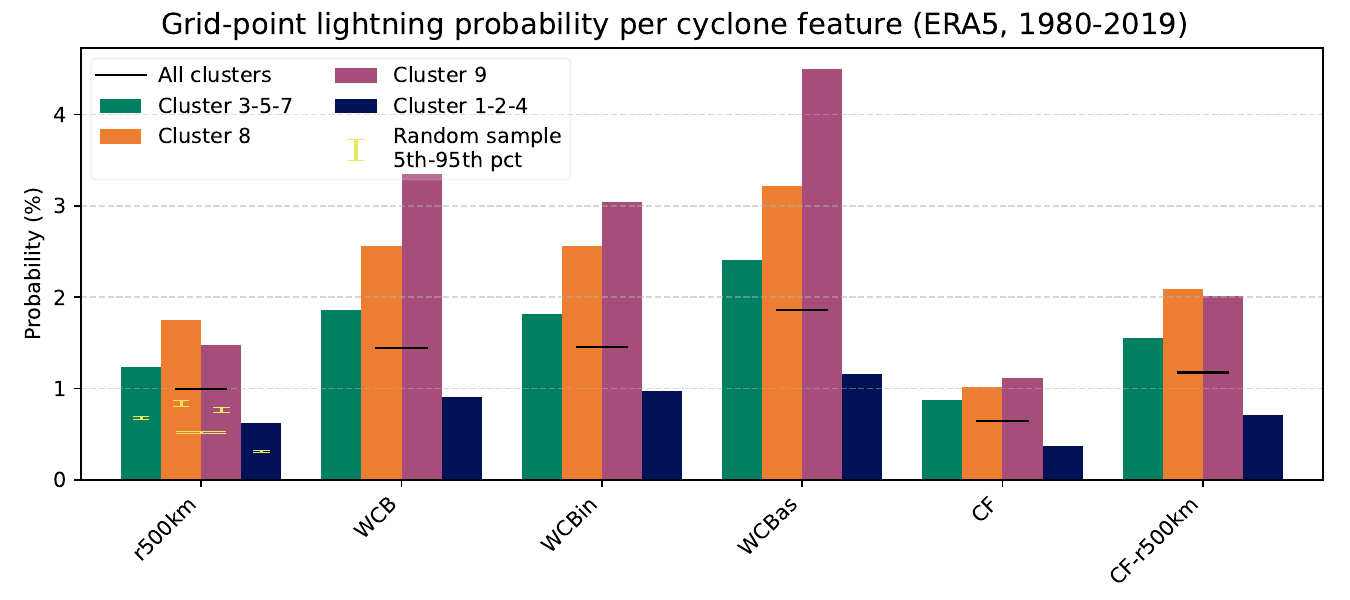}
\caption{Average lightning probability within different cyclone features, or combinations of co-occurring and overlapping features, per cyclone cluster group (coloured bars) and aggregated over all clusters (black horizontal line). All the cyclone time steps are considered. The yellow error bars show the 5th and 95th percentiles of lightning probability averages computed from 2000 Monte-Carlo samples of 500~km-radius circles (see Section~\ref{sec:met})}
\label{fig:light-feat}
\end{figure}

The composites in Figure~\ref{fig:composites} give a first indication of the link between convective activity and WCBs and CFs. This link is further explored by calculating the AR\-CHaMo lightning probability within different MedCy dynamical features, namely in the 500~km radius around the cyclone centre (r500km)\footnote{This central radius is equivalent to that used for the impact area. Its size is comparable with CF and WCB objects.}, in the WCB -- distinguished into WCB inflow (WCBin) and WCB ascent (WCBas) region, and in CFs. The WCB and CF objects always intercept the 500~km radius of a MedCy, and often extend outside (CFs especially). Dry intrusions, although known for enhancing CP in preceding CFs \citep{catto2019climatology,raveh2019climatology}, are neglected because of the weak lightning activity compared to climatologicy (not shown).

The black horizontal lines in Figure~\ref{fig:light-feat} indicate a grid-point (area-weighed) average probability per feature, computed across all cyclone time steps in the common dataset period 1980--2019. The lightning probability is $\sim$1\% within the r500km and increases to $\sim$1.4\% in the WCB, where it is higher in the WCBas compared to the WCBin region. 
We note that WCB objects, devised for Atlantic cyclones \citep[see][]{madonna2014warm}, capture the strongest warm-sector ascent around MedCys \citep[e.g.,][]{raveh2016large}.
In CFs the lightning probability is $\sim$0.6\%, but almost doubles ($\sim$1.2\%) when considering the CF regions within r500km. 
All values exceed the climatology (yellow errorbars in Figure~\ref{fig:light-feat}).

The coloured bars in Figure~\ref{fig:light-feat} show the lightning probabilities by cluster group. Despite the inter-group variability, feature-dependent changes are consistent with the description above. In Cl~9 WCBs amplify the lightning activity more strongly compared to other clusters. This is consistent with the fact that within Cl~9 WCBs produce lift in a region characterised by climatologically high CAPE (Figure~\ref{fig:composites}(g)), and thus may trigger DMC more often than in other clusters, producing the amplified increase in lightning frequency.

In Figure~S3 the feature-centred AR\-CHaMo ERA5 lightning probabilities are compared with ATDnet lightning-hour frequencies.
Their similarity (upon re-scaling) confirms the role played by the WCBs in supporting DMC around MedCys.

\section{Conclusions}
In this work we investigate convection around Mediterranean cyclones (MedCys) from a climatological perspective. Using observations and reanalysis, we distinguish the presence of deep, moist convection (DMC) and severe convective environments (SCEs), and discuss the presence of weak convection producing precipitation.

A preliminary statistical analysis on lightning observations shows that in the Mediterranean region DMC occurs mainly over or near the sea in autumn and winter, and over land in spring and summer. In autumn, winter and spring 20\% to 60\% of lightning over the sea and coastal areas is associated with MedCys. Furthermore,the presence of MedCys more than doubles the lightning frequency.

We answer the questions listed in the \nameref{sec:intro}.
\begin{enumerate}
    \item The warm sectors of MedCys favour the presence of SCEs because of low-level convergence ahead of the cold fronts, high thermodynamic instability and large-scale ascent. SCEs are frequent in cyclone clusters with peak occurrence in the transition seasons and summer (No.~3,~5,~7,~8,~9).
    The cluster with most pronounced convective characteristics (No.~8) differs from the other transition-season ones (No.~3,~5,~7) in terms of an autumn peak, a frequent occurrence in the northern Mediterranean, and a deep PV anomaly (see Figure~4 in \cite{givon2024process}). The low-level PV in cluster No.~8 is likely linked to the release of latent heat by convective processes.
    
    Winter baroclinic clusters (No.~1,~2,~4), although characterised by weak thermodynamic instability, display the largest accumulations of convective precipitation. We hypothesise that weak warm-sector CAPE environments, combined with lift provided by warm conveyor belts and by orography, trigger shallow or elevated convection producing precipitation \citep{oertel2019convective,oertel2021observations,fuhrer2005embedded}. Weak (above-zero) CAPE is favoured by the frequent propagation of the winter clusters over the warm Mediterranean Sea.

    \item The maximum frequency of convective environments and hazards (convective precipitation, lightning and hail) is located north-east of MedCy centres and in their warm (south-eastern) sectors. Confirming previous literature \citep{galanaki2016lightning,flaounas2018heavy}, environmental convective parameters and associated hazards are highest 12 to 6 hours before the time of maximum MedCy intensity.
    
    From a feature-centred perspective, the ascending branch of MedCys' warm conveyor belts shows a high modelled lightning probability, while cold fronts, known to favour convection by forcing low-level convergence \citep[e.g.,][]{naud2015cloudsat,rasp2016convective}, are associated with lower lightning probabilities. Although these findings are coherent with studies focusing on the United States \citep{jeyaratnam2020upright}, a geographical extension of our results is necessary to assert their validity for extratropical cyclones in general.
\end{enumerate}

The study of the relationship between MedCys and convection becomes increasingly relevant in view of recent and expected increases in Mediterranean SSTs \citep{pastor2020warming,soto2020evolution}. 
Understanding how different categories of MedCys are associated with SCEs is crucial for severe-weather risk assessment in operational forecasting. For this purpose, we recommend the detection of high CAPE environments useful to identify severe convective potential around cyclones. The evolution of the cyclones' trajectories is an additional factor influencing the persistence of the related hazards. This information is necessary to issue targeted and timely alerts and to manage meteo-hydrological emergencies. On top of this, statistical data at longer time scales is valuable for the insurance industry.

\bibliography{bibl_Med_01}

\end{document}